\documentclass[%
 reprint,
 amsmath,amssymb,
prb,
]{revtex4-1}
\usepackage{graphicx}% Include figure files
\usepackage{dcolumn}% Align table columns on decimal point
\usepackage{bm}% bold math
\usepackage[mathlines]{lineno}% Enable numbering of text and display math
\begin{document}

\preprint{APS/123-QED}

\title{Spin Magnetic Moment and Persistent Orbital Currents in Cylindrical Nanolayer}% Force line breaks with \\

%1, 1,

\author{N.G. Aghekyan}
\author{S.M. Amirkhanyan}
\author{E.M. Kazaryan}
\author{H.A. Sarkisyan}
\altaffiliation[Also at ]{Yerevan State University, 0025 Al. Manukyan, Yerevan, Armenia}
\email{shayk@ysu.com}
\affiliation{ Russian-Armenian (Slavonic) University, 123 H. Emin, Yerevan, Armenia }


%\affiliation{
%\affiliation{
%\affiliation{%

\date{\today}% It is always \today, today,

\begin{abstract}
Densities of persistent orbital and spin magnetic moment currents of an electron in a cylindrical nanolayer in the presence of external axial magnetic field are considered. For the mentioned current densities analytical expressions are obtained. The conditions when in the system only spin magnetic moment current is present are defined. Dependencies of orbital and spin magnetic moment currents on geometrical parameters of nanolayer are derived. It is shown that in the case of layered geometry the dependence of spin magnetic moment current on radial coordinate has a non-monotonic behavior. This is the peculiarity of layered geometry of nanostructure and it is due to the behavior of wave function of the system along the radial direction. Dependent on the directions of the field and orbital rotation of the electron there are defined values of radial coordinates when the orbital current disappears. The transition to the case of cylindrical quantum dot is discussed as well.
%\begin{description}
%\end{description}
\end{abstract}

\pacs{Valid PACS appear here}% PACS, the Physics and Astronomy
\maketitle

%\tableofcontents

\section{Introduction}

The development of precise methods of the growth of semiconductor nanostructures makes possible the realization of zero-dimensional systems of various geometrical forms and sizes \cite{Bimberg}. As we know quantum dots (QD) are those structures where quantum effects visualize themselves more vividly. Indeed, due to the full quantization of the spectrum of charge carriers, QDs have similar properties to real atoms, hence, usually, QDs are called "artificial atoms" \cite{Maksym_Chakraborty}. They are considered to be very perspective systems that can be used as base elements for contemporary nanoelectronics: started from lasers based on QDs and finished with new generation solar cells \cite{Asryan_Luri_2001, Asryan_Suris_1996, Nozik_2002, Aroutiounian_Petrosyan_2001, Cao_Yoon_2007}. The physical processes in QDs are being investigated intensively by the specialists, because besides the pure academic interest, the results of investigation can be applicable \cite{Alferov}. Currently spherical, cylindrical, ellipsoidal, pyramidal, lens-shaped and many other QDs are experimentally realized and thoroughly investigated \cite{Ozmen_Cakir, Safar_Barati_2013,Chen_Xie_2013,Safar_Barati_2012,Antonov_Daniltsev_2012,Soylu_2012,Zhang_Han_2012,Munacute_Domin_2012}. All of these QDs listed above have specific geometries that impose their electronic, optical, kinetic and other characteristics. The latest circumstance gives a vast choice of properties of QDs for solving a concrete practical problem.
An interesting class of QDs are layered QDs, for which there are two boundaries of transition from QD to the surrounding medium. This makes possible to control the physical properties of layered samples easily. In many papers it was considered electronic, optical and also thermal characteristics of ring-like and other layered nanostructures. The theoretical investigation of electron states in layered nanostructures is originated from the pioneering works of Chakraborty and Pietilainen \cite{Chak_Piet_1993,Chak_Piet_1994}. Authors have considered one-electron and many-electron states in quantum rings at the presence of impurities, as well as under the influence of a magnetic field. At the same time, taking into account that in the radial direction the movement of electron is restricted both on internal and external radiuses, Chakraborty and Pietilainen have suggested a model of confining potential having the form of a two-dimensional shifted oscillator. In Ref. \onlinecite{Aghek_Kaz_2010,Aghek_Kaz_Kost_2011,Aghek_Kaz_2012} the authors have discussed one and two electronic states in spherical nanolayer with different confinement potentials. Energy spectrum and wave functions have been obtained dependent on inner and outer radiuses. Along with changing inner or outer radiuses it is also possible to control physical properties of nanolayers with external electrical and magnetic fields, with hydrostatic pressure and so on. This means that it is urgent to study such systems by investigating interband and intraband absorption coefficients, ballistic conductance of orbital and spin currents and etc. Particularly the problem of charge current in ring like structures was discussed in many papers. For example in Ref. \onlinecite{Chak_Piet_1994} there have been studied the effect of electron-electron interaction on the magnetic moment (associated with the persistent current) of electrons in a quantum ring. There was introduced a model where the electron makes a circular motion in a parabolic confinement simulating a quantum ring which is subjected to a perpendicular magnetic field. The electron states in such a ring with and without the Coulomb interaction are then investigated. There also explored the limits of narrow and wide rings. In Ref. \onlinecite{Nita_Marinescu_2012} it was demonstrated the theoretical possibility of obtaining a pure spin current in a 1D ring with spin-orbit interaction by irradiation with a non-adiabatic, two-component terahertz laser pulse, whose spatial asymmetry is reflected by an internal phase difference. In Ref. \onlinecite{Castelano_2008} the persistent current in two vertically coupled quantum rings containing few electrons is studied. It was shown that the Coulomb interaction between the rings in the absence of tunneling affects the persistent current in each ring and the ground-state configurations. Quantum tunneling between the rings alters significantly the ground state and the persistent current in the system. Also this problem is discussed in Refs. \onlinecite{Chak_Piet_1993,Nita_Marinescu_2011,Bellucci_2009}.

In general, the quantum mechanical expression for one electron current in presence of magnetic field with consideration of the spin of electron consists of two components. The first one characterizes the orbital current ${\overrightarrow j _{Orb}}$ and it is connected with orbital motion of the electron. The second one is caused by the magnetic moment of electron and it is called density of spin magnetic moment current - ${\overrightarrow j_{SMM}}$. As far as this current is not conditioned with directed motion of the electron, therefore, its divergence equals to zero:

$$div{\overrightarrow j_{SMM}} = 0.$$

This current is specific and its existence is and is caused by the presence of the spin magnetic moment of the electron. In Ref. \onlinecite{Mita_Bougaida_1999} the peculiarities of spin magnetic moment current were discussed, particularly, when hydrogen atom electron is in  and states. In these states, the total current is formed exclusively by the spin magnetic moment current of the electron. It is interesting to note, that similar problem arises also in optics. Particularly, in Ref. \onlinecite{Berry_2009} it is shown that for scalar light, the current is the familiar expectation value of the intensity-weighted momentum operator. Current is distinct from the local wave vector (not weighted) that could be observed by means of quantum weak measurement. For vector light, the current (Poynting vector) contains an additional term corresponding to the photon spin, recently identified for paraxial light by Bekshaev and Soskin but valid generally after a modification to restore electric-magnetic democracy this term has physical consequences. Also in Ref. \onlinecite{Bakshaev_Bliokh_2011} it was discussed an optical phenomena associated with the internal energy redistribution which accompany propagation and transformations of monochromatic light fields in homogeneous media. The total energy flow (linear-momentum density, Poynting vector) can be divided into a spin part associated with the polarization and an orbital part associated with the spatial inhomogeneity. It is clear that in ring-like nanostructures it is also possible to discuss density of one electron current in presence of magnetic field with considering the spin. It is important to note, that in nanostructures there could be controlled densities of orbital as well as spin magnetic moment currents by the variation of geometrical parameters of studied samples.

In this paper the investigation of orbital and spin magnetic moment current densities for an electron located in cylindrical nanolayer are presented in presence of axial magnetic field.

\section{Cylindrical nanolayer}

\subsection{Energy spectrum and wave functions}

\begin{figure}
\begin{center}
\includegraphics[scale=0.6]{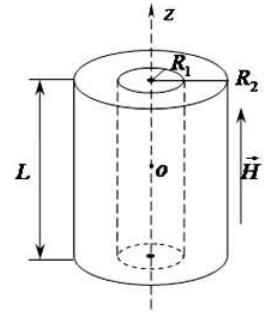}
\end{center}
\caption{\label{fig1} Geometrical shape of cylindrical nanolayer with $R_1$ inner and $R_2$ outer radiuses and $L$ height.}
\end{figure}

Let us discuss the behavior of electron in the cylindrical nanolayer with confinement potential

\begin{equation}
\label{eq1}
{V_{Conf}}\left( {\rho ,z} \right) = \left\{ \begin{array}{l}
0,\,\,{R_1} < \rho  < {R_2},\,\,\left| z \right| < \frac{L}{2}\\
\infty ,\,\,\rho  \le {R_1},\,\,\rho  \ge {R_2},\,\,\left| z \right| \ge \frac{L}{2}
\end{array} \right.,
\end{equation}

\noindent
where $L$ is the height of the cylindrical nanolayer,  $R_1$ and $R_2$ are respectively the inner and outer radiuses (Figure \ref{fig1}).

Let us also consider that the system is in an axial $\overrightarrow H $ homogenous magnetic field with the following gauge:

\begin{equation}
\label{eq2}
\overrightarrow A  = \left\{ {{A_\rho } = {A_z} = 0,{A_\varphi } = \frac{{H\rho }}{2}} \right\}.
\end{equation}

\noindent
For the gauge chosen above the variables in Schrodinger equation can be separated. By taking into account the spin of the electron one can obtain the following three dimensional equation:

\begin{widetext}
\begin{equation}
\label{eq3}
\begin{array}{l}
 - \frac{{{\hbar ^2}}}{{2\mu }}\left( {\frac{1}{\rho }\frac{\partial }{{\partial \rho }}\left( {\rho \frac{\partial }{{\partial \rho }}} \right) + \frac{{{\partial ^2}}}{{\partial {z^2}}} + \frac{1}{{{\rho ^2}}}\frac{{{\partial ^2}}}{{\partial {\varphi ^2}}} - \frac{{i\hbar {\omega _H}}}{2}\frac{\partial }{{\partial \varphi }} + \frac{{\mu \omega _H^2}}{8}{\rho ^2} + {V_{Conf}}\left( {\rho ,z} \right) - {\mu ^ * }H{{\hat \sigma }_z}} \right)\psi \left( {\rho ,\varphi } \right)f\left( z \right){\chi _{{s_z}}} = \\
 = {E_{{n_z},{n_\rho },m,{s_z}}}\psi \left( {\rho ,\varphi } \right)f\left( z \right){\chi _{{s_z}}}
\end{array}
\end{equation}
\end{widetext}

\noindent
with the following boundary conditions:

\begin{equation}
\label{eq4}
\Psi {|_{z =  \pm \frac{L}{2}}} = \Psi {|_{\rho  = {R_1}}} = \Psi {|_{\rho  = {R_2}}} = 0
\end{equation}

\noindent
where $\mu^*$ is the effective magnetic moment of electron (for $GaAs$ ${\mu ^ * } = {g_L}{\mu _B}$, ${\mu _B} = \frac{{e\hbar }}{{2{m_e}c}}$, ${g_L} = 0,44$, ${\hat \sigma _z}$ is the $z$ component of Pauli matrices, ${\omega _H} = \frac{{eH}}{{\mu c}}$ is the cyclotron frequency, where $\mu  = 0,067{m_e}$, $m_e$ electron rest mass and ${\chi _s}$ is the spin part of the wave function. For the   $z$ representation of Pauli matrices one can write:

\begin{equation}
\label{eq5}
{\hat \sigma _z} = \left( {\begin{array}{*{20}{c}}
1&0\\
0&{ - 1}
\end{array}} \right),\,\,{\chi _{{s_z} = \frac{1}{2}}} = \left( \begin{array}{l}
1\\
0
\end{array} \right)
\end{equation}

\noindent
Equation (\ref{eq3}) has analytical solutions. After corresponding calculations for the wave function one can obtain the following expression:

\begin{equation}
\label{eq6}
\begin{split}
\begin{array}{l}
\Psi _{{n_\rho },m}^{n,\,{s_z}}\left( {\rho ,\varphi ,z,{s_z}} \right) = \frac{1}{{\sqrt {2\pi } }}{e^{im\varphi }}\sqrt {\frac{2}{L}} \cdot   \\ \cdot \left\{ \begin{array}{l}
\sin \frac{{\pi n}}{L}z\,(n = 2k)\\
\cos \frac{{\pi n}}{L}z\,(n = 2k + 1)
\end{array} \right\}\left( \begin{array}{l}
1\\
0
\end{array} \right)R\left( \rho  \right)
\end{array}
\end{split}
\end{equation}

\noindent
where $m = 0; \pm 1; \pm 2;...$ is the magnetic quantum number.

After substitution (\ref{eq6}) into (\ref{eq3}) we obtain an equation for the radial wave function.After inserting the following notations:

$$\beta  = \frac{{E - {E_n}}}{{\hbar {\omega _H}}} - \frac{m}{2},\,\,\\ {a_H} = \sqrt {\frac{\hbar }{{{\omega _H}\mu }}} \text{ - magnetic length}$$

\noindent
the solutions for that radial equation can be written as follows:

\begin{widetext}
\begin{equation}
\label{eq7}
R\left( \rho  \right) = {\rho ^{\left| m \right|}}{e^{ - \frac{{{\rho ^2}}}{{4a_H^2}}}}\left\{ {{C_1} \cdot {}_1{F_1}\left( { - \left( {\beta  - \frac{{\left| m \right| + 1}}{2}} \right),\,m + 1,\,\frac{{{\rho ^2}}}{{2a_H^2}}} \right) + {C_2} \cdot U\left( { - \left( {\beta  - \frac{{\left| m \right| + 1}}{2}} \right),\,m + 1,\,\frac{{{\rho ^2}}}{{2a_H^2}}} \right)} \right\}
\end{equation}
\end{widetext}

\noindent
where $_1{F_1}$ and $U$ are Kummer confluent hypergeometric and confluent hypergeometric functions correspondingly. On Figure \ref{fig2} the distribution of electron probability density is depicted. As it follows from Figure \ref{fig2} the maximal probability of localizing of electron is in the center of the nanolayer in radial direction.
In order to find energy spectrum for the ground state, it is necessary to consider the radial boundary conditions (\ref{eq4}) and solve the obtained transcendental equation below:

\begin{figure}
\begin{center}
\includegraphics[scale=0.6]{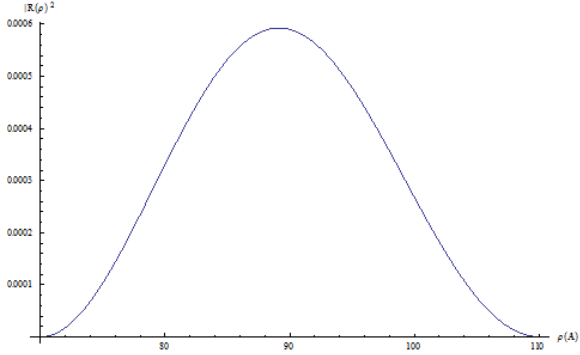}
\end{center}
\caption{\label{fig2} Distribution of density of probability of electron, when $ H  = {10^4}{g^{\frac{1}{2}}}\cdot{m^{ - \frac{1}{2}}}\cdot{s^{ - 1}}$}
\end{figure}

\begin{widetext}
\begin{equation}
\label{eq8}
\left| \begin{array}{l}
_1{F_1}\left( { - \left( {\beta  - \frac{{\left| m \right| + 1}}{2}} \right),\,m + 1,\,\frac{{R_1^2}}{{2a_H^2}}} \right)\,\,\,\,\,U\left( { - \left( {\beta  - \frac{{\left| m \right| + 1}}{2}} \right),\,m + 1,\,\frac{{R_1^2}}{{2a_H^2}}} \right)\\
_1{F_1}\left( { - \left( {\beta  - \frac{{\left| m \right| + 1}}{2}} \right),\,m + 1,\,\frac{{R_2^2}}{{2a_H^2}}} \right)\,\,\,\,\,U\left( { - \left( {\beta  - \frac{{\left| m \right| + 1}}{2}} \right),\,m + 1,\,\frac{{R_2^2}}{{2a_H^2}}} \right)
\end{array} \right| = 0
\end{equation}
\end{widetext}

\noindent
As a result of those operations we will get:

\begin{equation}
\label{eq9}
{E_{n,m}} = \hbar {\omega _H}\left( {\beta  + \frac{m}{2}} \right) + {E_n}
\end{equation}

\noindent
where $\beta$ can be obtained by numerical calculations.

\subsection{Orbital current}
On the basis of the results obtained above one can calculate orbital and spin magnetic moment current densities for the discussed system:

\begin{equation}
\label{eq10}
{\vec j_{Tot}} = {\vec j_{Orb}} + {\vec j_{SMM}}
\end{equation}

\noindent
where the first term represents the orbital and the last one - spin magnetic moment current.

According to the theory of quantum mechanics the expression of orbital current density of a charge carrier in the magnetic field has the following form:

\begin{equation}
\label{eq11}
{\vec j_{Orb}} = \frac{{ie\hbar }}{{2\mu }}\left( {\Psi \vec \nabla {\Psi ^ * } - {\Psi ^ * }\vec \nabla \Psi } \right) - \frac{{{e^2}}}{{\mu c}}\vec A{\left| \Psi  \right|^2}
\end{equation}

\noindent
and the spin magnetic moment current density as follows \cite{Landau_Lifshitz_2003}:

\begin{equation}
\label{eq12}
{\vec j_{SMM}} = {\mu ^ * }c\,rot\left( {{\Psi ^ * }\hat \vec \sigma \,\Psi } \right).
\end{equation}

By direct calculations it could be shown that

\begin{equation}
\label{eq13}
\begin{array}{l}
{\left( {{{\vec j}_{Orb}}} \right)_\rho } = 0,\,\,\,{\left( {{{\vec j}_{Orb}}} \right)_z} = 0,\,\,\\
\,{\left( {{{\vec j}_{Orb}}} \right)_\varphi } = \left( {\frac{{e\hbar m}}{{\mu \rho }} - \frac{{{e^2}H}}{{2\mu c}}\rho } \right){\left| {\psi (\rho ,\varphi )} \right|^2}{\left| {f(z)} \right|^2},
\end{array}
\end{equation}

\begingroup
\squeezetable
\begin{table*}
\caption{}
\label{table1}
\begin{center}
For $m = 1$ \\
\begin{tabular}{ |c|c|c|c|c|c| }
  \hline
  $\left| {\overrightarrow H } \right|({g^{\frac{1}{2}}} \cdot c{m^{\frac{1}{2}}} \cdot {s^{ - 2}})$ & $-1.1 \cdot 10^5$ & $-1.5 \cdot 10^5$ & $-1.9 \cdot 10^5$ & $-2.3 \cdot 10^5$ & $-2.65 \cdot 10^5$ \\ \hline
  ${\rho _m}(\mathop {\rm A}\limits^ \circ  )$ & 109.43 & 93.71 & 83.26 & 75.68 & 70.50 \\
  \hline
\end{tabular}
\end{center}
\begin{center}
~\newline
~\newline
For $\left| {\overrightarrow H } \right| = 2.65*{10^5}{g^{\frac{1}{2}}}*c{m^{\frac{1}{2}}}*{s^{ - 2}}$ \\
\begin{tabular}{ |c|c|c|c|c|c| }
  \hline
  $m$ & 1 & 2 & 3 & 4 & 5 \\ \hline
  ${\rho _m}(\mathop {\rm A}\limits^ \circ  )$ & 70.50 & 99.71 & 122.12 & 141.01 & 157.65 \\
  \hline
\end{tabular}
\end{center}
\end{table*}
\endgroup

\noindent
where $\psi (\rho ,\varphi )$ and $f(z)$ are the two components of $\Psi $ final wave function dependent on correspondingly $\rho ,\,\varphi $ and $z$ variables. The orbital current has two components. If we consider that $\hbar m$ is the orbital momentum characterizing the state with quantum number $m$, then it is clear that the first term in formula (\ref{eq13}) characterizes the current conditioned by the orbital motion. In its turn $\frac{{eH}}{{\mu c}}$ is the cyclotron frequency of the electron. The multiplication of cyclotron frequency and $\rho $ characterizes the linear velocity of cyclotron movement of the electron. Therefore, the second term is the cyclotron frequency current and describes the contribution of the magnetic field in ${\vec j_{Orb}}$.

As we can see from (\ref{eq13}), when $\frac{{e\hbar m}}{{\mu \rho }} = \frac{{{e^2}H}}{{2\mu c}}\rho $ correspondingly ${\rho _m} = \sqrt {2m} {a_H}$ and when radial coordinate of the electron has this value the orbital current in nanolayer becomes zero. Let us discuss it in details. As far as we consider electron so let us write this equation as $ - \frac{{\left| e \right|\hbar m}}{{\mu \rho }} = \frac{{{e^2}H}}{{2\mu c}}\rho $. For this equation one of the following two conditions should be satisfied:

\begin{enumerate}
  \item $m > 0$ and $\overrightarrow H  \uparrow  \downarrow Oz$, which means that $\overrightarrow H $ has the opposite direction with respect to $Oz$ axis
  \item $m < 0$	and $\overrightarrow H  \uparrow  \uparrow Oz$, which, respectively, means that  has the same direction as $Oz$ axis
\end{enumerate}

It is important to notice that for each fixed value of $m$ there is its own fixed value of ${\rho _m}$ . For a given value of magnetic field $\left| {\overrightarrow H } \right|$ on distances equal to ${\rho _m} = \sqrt {2m} {a_H}$ orbital current equals to zero, as it was mentioned later, and when the radialcoordinate of the electron is larger than this value we have oppositelydirected orbital current. Physically this phenomenon could be explained as follows. With increasing $\rho $ the contribution of energy caused by angular momentum decreases but at the same time contribution of energy of cyclotron rotation increases. For the radiuses larger than ${\rho _m}$ the direction of rotation is mainly conditioned by magnetic field $\overrightarrow H $. On Table \ref{table1} there are shown several values of ${\rho _m}$ for different values of $\overrightarrow H $ and $m$.

In order to calculate orbital and spin magnetic moment current densities, first we should find integration constants, therefore, we should use the boundary conditions (\ref{eq4}). Thus, for those coefficients we obtain the following expressions:

\begin{equation}
\label{eq14}
{C_2} =  - \frac{{{}_1{F_1}\left( { - \left( {\beta  - \frac{{\left| m \right| + 1}}{2}} \right),\,m + 1,\,\frac{{{R_1}^2}}{{2a_H^2}}} \right)}}{{U\left( { - \left( {\beta  - \frac{{\left| m \right| + 1}}{2}} \right),\,m + 1,\,\frac{{{R_1}^2}}{{2a_H^2}}} \right)}}{C_1}
\end{equation}

\noindent
from where, by using the normalization condition we find, that

\begin{figure}
\begin{center}
\includegraphics[scale=0.7]{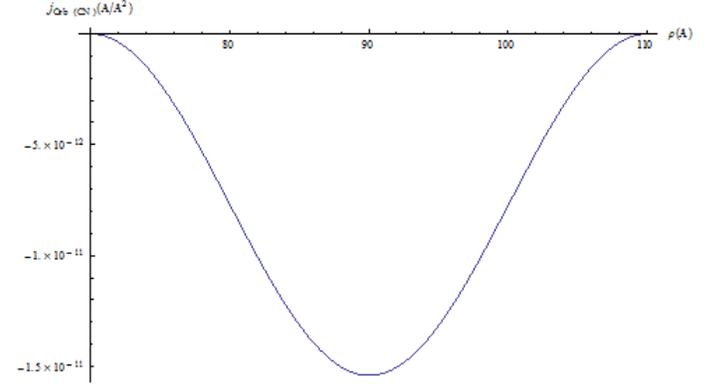}
\end{center}
\caption{\label{fig3} Dependence of orbital current density on radial coordinate for quantum cylindrical nanolayer, when $ H  = {10^4}{g^{\frac{1}{2}}}\cdot{m^{ - \frac{1}{2}}}\cdot{s^{ - 1}}$}
\end{figure}

\begin{widetext}
\begin{equation}
\label{eq15}
{C_1} = \frac{1}{{\sqrt {\int\limits_{{R_1}}^{{R_2}} {{\left( {{\rho ^{\left| m \right|}}{e^{^{ - \frac{{{\rho ^2}}}{{4a_H^2}}}}}}  {\left[{}_1{F_1}\left( {\frac{{\left| m \right| + 1}}{2} - \beta ,\,m + 1,\,\frac{{{\rho ^2}}}{{2a_H^2}}} \right) - \frac{{{}_1{F_1}\left( {\frac{{\left| m \right| + 1}}{2} - \beta ,\,m + 1,\,\frac{{{R_1}^2}}{{2a_H^2}}} \right)}}{{U\left( {\frac{{\left| m \right| + 1}}{2} - \beta ,\,m + 1,\,\frac{{{R_1}^2}}{{2a_H^2}}} \right)}}U\left( {\frac{{\left| m \right| + 1}}{2} - \beta ,\,m + 1,\,\frac{{{\rho ^2}}}{{2a_H^2}}} \right)\right]} \right)}^2}\rho d\rho } }}
\end{equation}
\end{widetext}

On Figure \ref{fig3} the dependence of orbital current density on radial coordinate for the ground state of electron ($n = 1, m = 0$) is illustrated. Let us discuss the behavior of these current densities in detail when $L = 500\text{\AA}$, $R_1 = 70\text{\AA}$, $R_2=110\text{\AA}$. In Figure \ref{fig3} it can be seen that the modulus of orbital current density fully repeats the radial distribution of the electron in the nanolayer and that is natural. As far as we consider the state when $m=0$, the orbital current is negative and its profile has the inverted form of density of probability of radial distribution.

\subsection{Spin magnetic moment current}

Now let us turn to the investigation of spin magnetic moment current. As for the spin magnetic moment current density calculation, first it is necessary to calculate the components of the following vector:

\begin{equation}
\label{eq16}
\bar \vec \sigma  = {\mu ^ * }{\Psi ^ * }\hat \vec \sigma \,\Psi
\end{equation}

\noindent
By taking into consideration the forms of Pauli matrices and ${\chi _s}$ wave function, with direct calculations it can be obtained that
\begin{equation}
\label{eq17}
{\left( {\bar \vec \sigma } \right)_\rho } = {\left( {\bar \vec \sigma } \right)_\varphi } = 0
\end{equation}

\noindent
and

\begin{equation}
\label{eq18}
{\left( {\bar \vec \sigma } \right)_z} = 2{\mu ^ * }{s_z}{\left| \psi  \right|^2}{\left| f \right|^2}.
\end{equation}

$\left( {{{\left| \psi  \right|}^2}{{\left| f \right|}^2}} \right)$ multiplication is only dependent on $\rho $ and $z$ coordinates, therefore different from zero will be only the $\varphi $ component of $rot\bar \vec \sigma $. For $\rho, \varphi$ and $z$ components of spin magnetic moment current density one can obtain:

\begin{equation}
\label{eq19}
\begin{gathered}
{\left( {{{\vec j}_{SMM}}} \right)_\rho } = 0,\,\,{\left( {{{\vec j}_{SMM}}} \right)_\varphi } =  - 2{\mu ^ * }c{s_z}\frac{\partial }{{\partial \rho }}{\left| \psi  \right|^2}{\left| f \right|^2},\\
{\left( {{{\vec j}_{SMM}}} \right)_z} = 0.
\end{gathered}
\end{equation}

\noindent
Finally, for the total current density it can be written:

\begin{equation}
\label{eq20}
\begin{gathered}
{\left( {{{\vec j}_{tot}}} \right)_\varphi } = \left( {\frac{{e\hbar m}}{{\mu \rho }}  - \frac{{{e^2}H}}{{2\mu c}}\rho } \right){\left| \psi  \right|^2}{\left| f \right|^2} -\\ - 2{\mu ^ * }c{s_z}\frac{\partial }{{\partial \rho }}{\left| \psi  \right|^2}{\left| f \right|^2}.
\end{gathered}
\end{equation}

These results are quite expected because in $\rho$  and $z$ directions charge carriers motion is limited. On the other hand we have periodical rotation of particle around $Oz$ axis which creates cyclical current.

It should be noted, that in contrast with the case of quantum wire, the consideration of the quantization along the $Oz$ axis leads to the vanishing of the orbital current in the direction of magnetic field. On the other hand both in orbital and in spin magnetic moment current expressions the presence of $Oz$ direction is expressed by the presence of ${\left| f \right|^2}$ factor. Therefore, dependent on which $z = const$ plain is discussed the current, for the same $\rho$ it will have different values. It is important to note that, in the expression of total current ${\left( {{{\vec j}_{tot}}} \right)_\varphi }$ the factor ${\left| f \right|^2}$ characterizes the quantization along $Oz$ axis. Herewith

\begin{equation}
\label{eq21}
{\left| f \right|^2} = \left\{ \begin{array}{l}
\frac{2}{L}{\sin ^2}\left( {\frac{{\pi n}}{L}z} \right)\left( {n - even} \right)\\
\frac{2}{L}{\cos ^2}\left( {\frac{{\pi n}}{L}z} \right)\left( {n - odd} \right)
\end{array} \right.
\end{equation}

\noindent
Hence, along $Oz$ axis of cylindrical nanolayer there are plains where total current equals to zero (see Figure \ref{fig4}). These plains match to ${\left| f \right|^2}$ factor zeros. Let us consider two cases: one for an even value of  and another for an odd one. For example, when $n=6$ we should consider ${\left| f \right|^2} = \frac{2}{L}{\sin ^2}\left( {\frac{{\pi n}}{L}z} \right)$.Then there will be seven flats where total current will be equal to zero. $z$ coordinates of these plains could be calculated from $\frac{2}{L}{\sin ^2}\left( {\frac{{6\pi }}{L}z} \right) = 0$ condition. For the case of $n=7$ (the odd case) zero plains are obtained by the same calculations for ${\left| f \right|^2} = \frac{2}{L}{\cos ^2}\left( {\frac{{\pi n}}{L}z} \right)$. These results could be illustrated as follows:

\begin{figure}
\begin{center}
\includegraphics[scale=0.6]{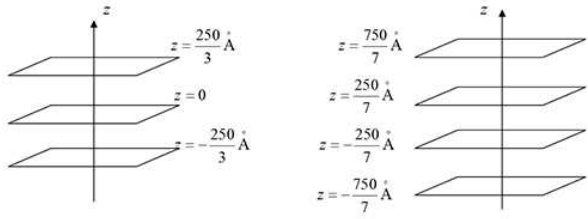}
\end{center}
\caption{\label{fig4} Illustration of plains where total current equals to zero.}
\end{figure}

One more important thing that should be noted: by fixing $m$ in expression (\ref{eq20}) one can choose the magnetic vector $\overrightarrow H $ so that in the total current density the contribution would have only the spin magnetic moment current:

\begin{equation}
\label{eq22}
{\left( {{{\vec j}_{tot}}} \right)_\varphi } =  - 2{\mu ^ * }c{s_z}{\left| f \right|^2}\left\{ {\frac{{\partial {{\left| \psi  \right|}^2}}}{{\partial \rho }}} \right\}.
\end{equation}

\noindent
which is conditioned by the gradient of electron radial distribution.

\begin{figure}
\begin{center}
\includegraphics[scale=0.6]{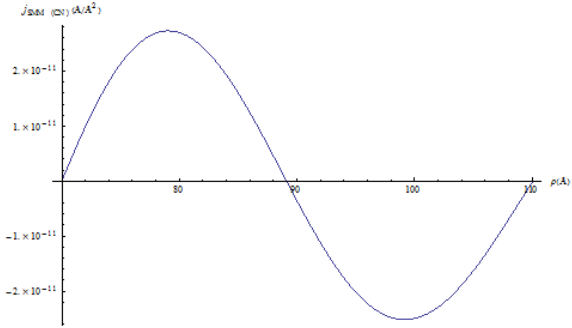}
\end{center}
\caption{\label{fig5} Dependence of spin magnetic moment current density on radial coordinate for quantum cylindrical nanolayer, when $ H  = {10^4}{g^{\frac{1}{2}}}\cdot {m^{ - \frac{1}{2}}}\cdot{s^{ - 1}}$}
\end{figure}

\begin{figure}
\begin{center}
\includegraphics[scale=0.6]{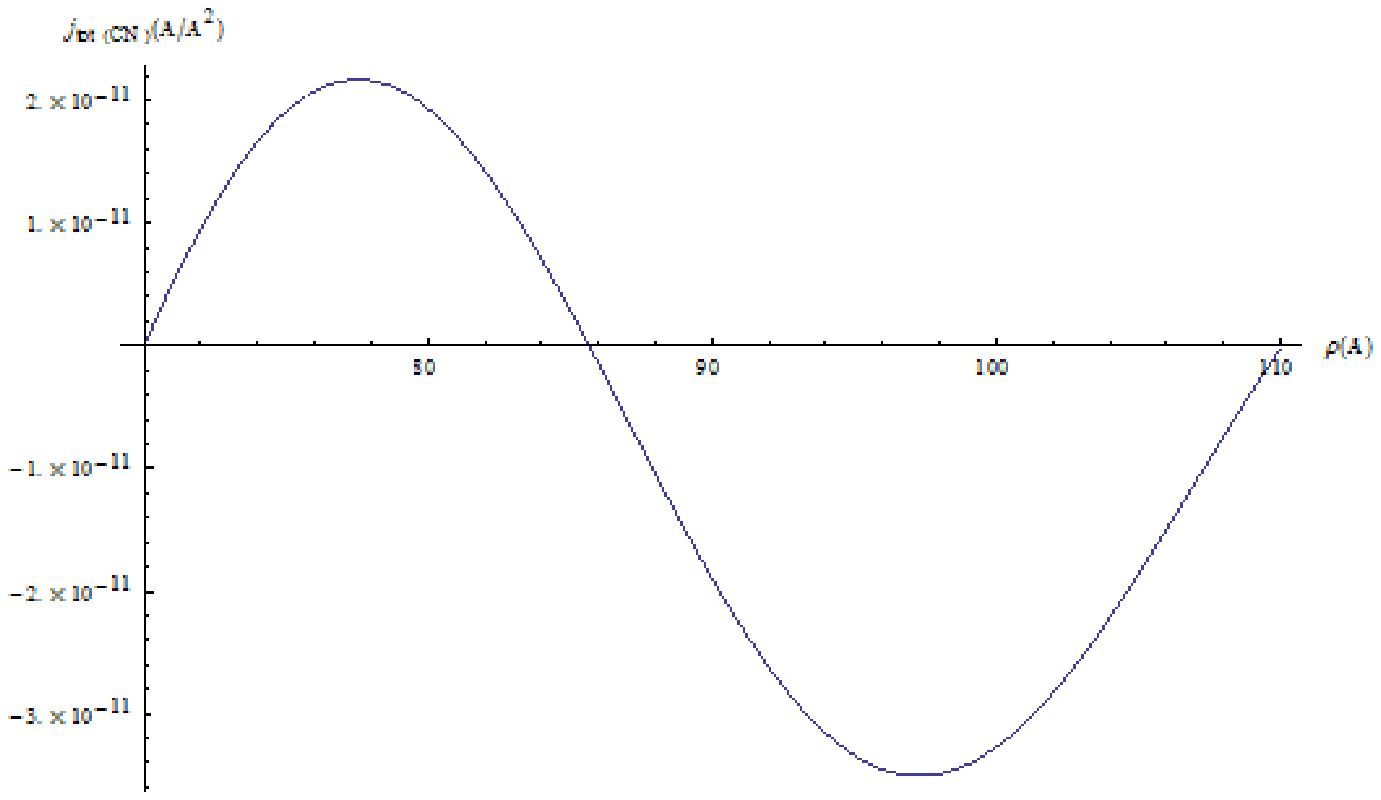}
\end{center}
\caption{\label{fig6} Dependence of total current density on radial coordinate for quantum cylindrical nanolayer, when $ H  = {10^4}{g^{\frac{1}{2}}}\cdot {m^{ - \frac{1}{2}}}\cdot{s^{ - 1}}$}
\end{figure}

\begin{figure}
\begin{center}
\includegraphics[scale=0.6]{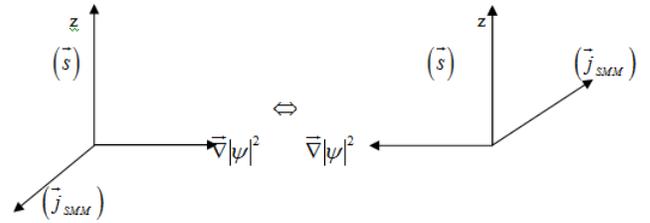}
\end{center}
\caption{\label{fig7} Right-handed orthogonal trio of the vectors.}
\end{figure}

On Figure \ref{fig5} and \ref{fig6} spin magnetic moment and total current densities are depicted respectively. As we can see on Figure \ref{fig5} spin magnetic moment has a non-monotonic character. Such behavior is specific for the layered geometries. As far as $\overrightarrow \nabla , \overrightarrow {{\mu ^ * }} $ and ${\overrightarrow j _{SMM}}$ form a right-handed orthogonal trio, from (\ref{eq12}) it follows that when $\overrightarrow \nabla  {\left| \psi  \right|^2}$ changes its direction to opposite ${\overrightarrow j _{SMM}}$ should change its direction to opposite too (see Figure \ref{fig7}).

In fact, the spin magnetic moment current is a pseudo current. Here under the spin magnetic moment current it must be understood imaginary currents which would create the same magnetic field distribution which is created by the spin magnetic moment of the electron. From physical point of view it is obvious that for the cylindrical nanolayer the spin magnetic moment should create a magnetic field parallel to the axis of the cylindrical nanolayer as the spin was directed by the axis originally. A magnetic field parallel to $Oz$ axis can be caused by oppositely directed currents in a plain perpendicular to the $Oz$ axis (Figure \ref{fig5}).

\section{Cylindrical quantum dot}

Let us also discuss the case of quantum cylinder, i. e. the limiting case when the inner radius of the cylindrical nanolayer is equal to zero. All the logic and calculations are the same as that for cylindrical nanolayer with only one difference: radial wave function has the following form \cite{Atayan_Kazaryan_2008}

\begin{equation}
\label{eq23}
R\left( \rho  \right) = {C_3}{\rho ^{\left| m \right|}}{e^{ - \frac{{{\rho ^2}}}{{4a_H^2}}}}{}_1{F_1}\left( {\frac{{\left| m \right| + 1}}{2} - \beta ,\,\left| m \right| + 1,\,\frac{{{\rho ^2}}}{{2a_H^2}}} \right)
\end{equation}

\noindent
where $\beta  = \frac{{{E_\rho }}}{{\hbar {\omega _H}}} + sign(e)\frac{m}{2}$. Again, by using the normalization condition we can find the integration constant and from boundary condition $R\left( {{\rho _0}} \right) = 0$, where $\rho_0$ is the cylinder radius, we can find the energy spectrum. Thereby, we obtain that \cite{Atayan_Kazaryan_2008}

$${\alpha _{{n_\rho } + 1,\left| m \right|}} = \beta  - \frac{{\left| m \right| + 1}}{2}$$

\noindent
where ${n_\rho } = 0;\,\,1;\,\,2;\,\,...$. Hence,

\begin{equation}
\label{eq24}
{E_{{n_\rho },m}} = \hbar {\omega _H}\left( {{\alpha _{{n_\rho } + 1,m}} - sign\left( e \right)\frac{m}{2} + \frac{{\left| m \right| + 1}}{2}} \right).
\end{equation}

\noindent
It should be mentioned, that these solutions are entirely analogous to the ones of non-diffracting Laguerre-Gaussian optical beams \cite{Bliokh_Schattschneider_2012}.

\begin{figure}
\begin{center}
\includegraphics[scale=0.7]{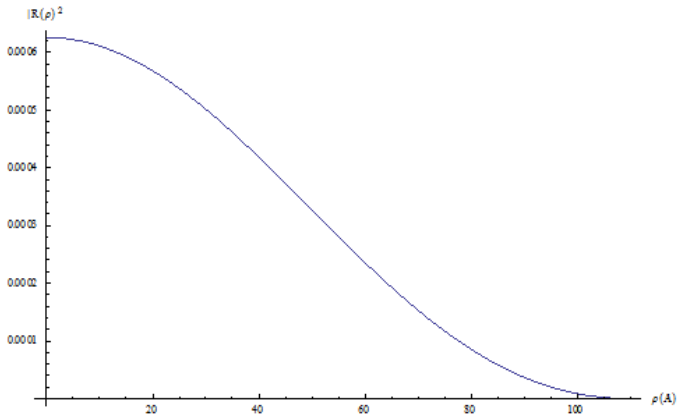}
\end{center}
\caption{\label{fig8} Distribution of electron density of probability for cylindrical QD, when $ H  = {10^4}{g^{\frac{1}{2}}}\cdot {m^{ - \frac{1}{2}}}\cdot{s^{ - 1}}$}
\end{figure}

On Figure \ref{fig8} the distribution of probability density of electron dependent on radial coordinate for the case of cylindrical quantum dot is depicted.

\begin{figure}
\begin{center}
\includegraphics[scale=0.6]{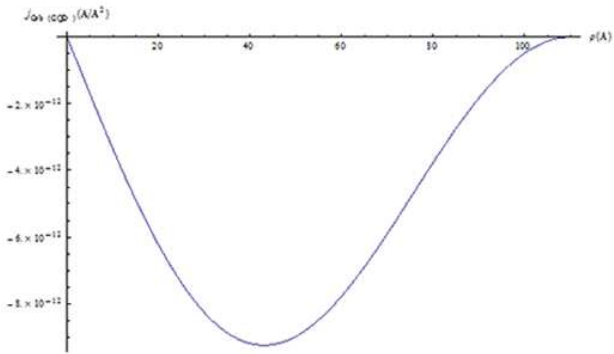}
\end{center}
\caption{\label{fig9} Dependence of orbital current density on radial coordinate for cylindrical QD, when $ H  = {10^4}{g^{\frac{1}{2}}}\cdot {m^{ - \frac{1}{2}}}\cdot{s^{ - 1}}$}
\end{figure}

On Figure \ref{fig9} it is shown the dependence of orbital current on radial coordinate. As we can see on periphery and in the center of the cylinder the current is equal to zero. On the periphery it equals to zero because wave function, and, therefore, the probability of electron localization there becomes zero. In the center of the system the current is equal to zero because there is no rotation of the electron.

\begin{figure}
\begin{center}
\includegraphics[scale=0.6]{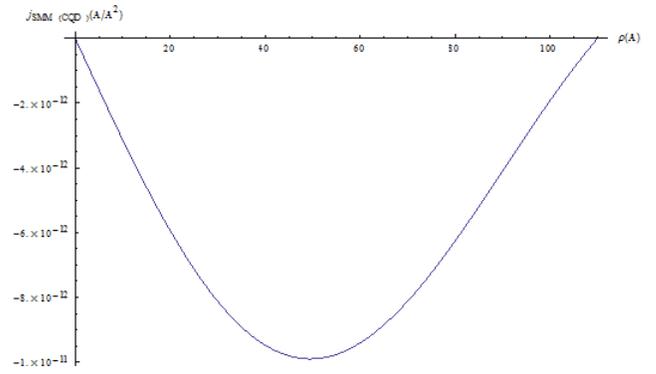}
\end{center}
\caption{\label{fig10} Dependence of spin magnetic moment current density on radial coordinate for cylindrical QD, when $ H  = {10^4}{g^{\frac{1}{2}}}\cdot {m^{ - \frac{1}{2}}}\cdot{s^{ - 1}}$}
\end{figure}

As for the spin magnetic moment current densities the behavior is shown on Figure \ref{fig10} from where we can see

\noindent
that we have current with spatial distribution of electron charge similar to the orbital current density. As in the case of cylindrical nanolayer we can say as far as we consider the state when $m=0$, the orbital current is negative and its profile has the inverted form of density of probability of radial distribution.

\begin{figure}
\begin{center}
\includegraphics[scale=0.7]{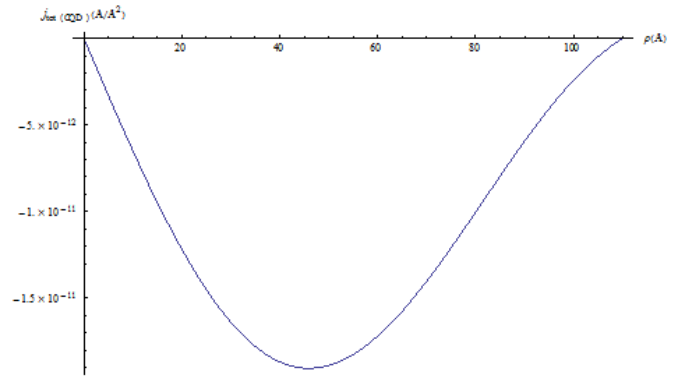}
\end{center}
\caption{\label{fig11} Dependence of total current on radial coordinate for cylindrical QD, when $ H  = {10^4}{g^{\frac{1}{2}}}\cdot {m^{ - \frac{1}{2}}}\cdot{s^{ - 1}}$}
\end{figure}

On Figure \ref{fig11} the dependence of total current on radial coordinate is illustrated.

\section{Conclusion}

Thereby, in this paper the orbital and spin magnetic moment currents of an electron in a cylindrical nanolayer are investigated. Analytical expressions are obtained for both of those currents. It was shown that under certain conditions the main contribution to the total current is due to the spin magnetic moment current. Particularly, there are cylindrical surfaces parallel to the axis of the nanolayer, where the orbital current becomes zero. It was shown that there are plains perpendicular to the axis of the nanolayer, where both the persistent orbital and spin magnetic moment currents become zero.

Also the calculations were done for the limiting case when the inner radius of the nanolayer is zero. It was illustrated that the spin magnetic moment current has non-monotonic behavior in the radial direction, whereas in the case of cylindrical QD it has monotonic character. Both for the cylindrical nanolayer and for cylindrical QD it was shown that the persistent orbital and spin magnetic moment current can be controlled by changing the geometrical parameters of the quantum nanostructures - the inner or outer radius in the case of the layer, and the radius in the case of the QD.

\section{Acknowledgements}

The authors thank Professor Konstantin Bliokh for useful discussions. The work was performed as a part of the basic program of state of the Republic of Armenia "Studies of the physical properties of quantum nanostructures with complex geometries and different limiting potentials".

\end{document}